# GTRSS: Graph-based Top-$k$ Representative Similar Subtrajectory Query

Mingchang Ge[†], Liping Wang[†], Xuemin Lin[‡], Yuang Zhang[†], Kunming Wang[§]
[†]*East China Normal University,* [‡]*Shanghai Jiao Tong University,* [§]*Fudan University*
51265902151,51255902045@stu.ecnu.edu.cn,
lipingwang@sei.ecnu.edu.cn, xuemin.lin@gmail.com, kmwamg24@m.fudan.edu.cn

*Abstract*—Trajectory mining has attracted significant attention. This paper addresses the Top-$k$ Representative Similar Subtrajectory Query (TRSSQ) problem, which aims to find the $k$ most representative subtrajectories similar to a query. Existing methods rely on costly filtering-validation frameworks, resulting in slow response times. Addressing this, we propose GTRSS, a novel Graph-based Top-$k$ Representative Similar Subtrajectory Query framework. During the offline phase, GTRSS builds a dual-layer graph index that clusters trajectories containing similar representative subtrajectories. In the online phase, it efficiently retrieves results by navigating the graph toward query-relevant clusters, bypassing full-dataset scanning and heavy computation. To support this, we introduce the Data Trajectory Similarity Metric (DTSM) to measure the most similar subtrajectory pair. We further combine R-tree and grid filtering with DTSM's pruning rules to speed up index building. To the best of our knowledge, GTRSS is the first graph-based solution for top-k subtrajectory search. Experiments on real datasets demonstrate that GTRSS significantly enhances both efficiency and accuracy, achieving a retrieval accuracy of over 90% and up to 2 orders of magnitude speedup in query performance.

## I. INTRODUCTION

Trajectory mining has wide applications, including traffic flow prediction [1], travel time estimation [2], and athlete behavior studies [3], where trajectory similarity is a fundamental and widely studied task. Current trajectory similarity studies have taken two distinct routes. The first focuses on metrics, like DTW [4], EDR [5], and ERP [6]. The second focuses on efficient search techniques, such as indexing strategy [7] and distributed method [8].

However, traditional trajectory similarity search may fail to capture meaningful matches due to differences in trajectory lengths. As shown in Fig. 1, the overall similarity between the query trajectory $T^q$ and the full data trajectories $T^{d_1}$ and $T^{d_2}$ may be low because of length discrepancies, even though certain segments of these trajectories may exhibit high similarity to the query. This highlights the need for a more nuanced approach—one that considers segments of trajectories rather than the whole trajectory.

Subtrajectory similarity analysis[9–12] addresses this by focusing on continuous segments of a trajectory that match the query, even if the full trajectory does not. For example, subtrajectories $T^{d_1}_{4:8}$ and $T^{d_2}_{2:8}$ in Fig. 1 are more similar to $T^q$ than their full trajectories. This demonstrates the advantage of subtrajectory similarity in providing more precise and relevant matches. Notable progress includes indexing and retrieval method [10], splitting-based algorithm [9], dynamic programming-based approach [13], and others. However, a single trajectory corresponds to multiple subtrajectories with similar patterns, the number of which is quadratic in the length of the full trajectory. Efficiently identifying representative similar subtrajectories remains a critical yet challenging task.

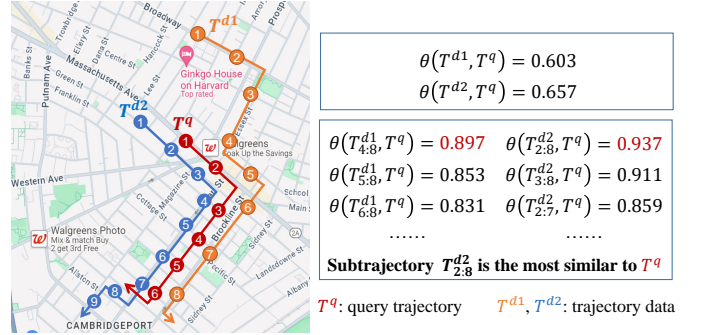

Fig 1: **An example of representative similar subtrajectory, For $T^q$, the representative similar subtrajectory in $T^{d_1}$ is $T^{d_1}_{4:8}$, and in $T^{d_2}$, it is $T^{d_2}_{2:8}$, with corresponding representative similarity scores of 0.897 and 0.937, respectively.**

**Example 1.** *as shown in Fig. 1, $T^{d_2}$ contains multiple subtrajectories such as $T^{d_2}_{2:8}$, $T^{d_2}_{3:8}$ and $T^{d_2}_{2:7}$. Subtrajectories $T^{d_2}_{2:8}$ and $T^{d_2}_{3:8}$ have similarity scores of 0.937 and 0.911 with $T^q$, respectively, which are the results of top-$k(k=2)$ subtrajectory similarity query. However, $T^{d_2}_{2:8}$ and $T^{d_2}_{3:8}$ differ by only a single point, bringing about a high redundancy that hinders the diversity of the query results. A more reasonable solution is to ensure that the results come from different trajectories. Each trajectory contributes only one representative subtrajectory most similar to query $T^q$, to participate in the top-k selection. For query $T^q$, the representative subtrajectories of $T^{d_1}$ and $T^{d_2}$ are $T^{d_1}_{4:8}$ and $T^{d_2}_{2:8}$, respectively. These two subtrajectories then constitute the top-$k(k=2)$ query results of representative similar subtrajectories.*

The query problem for **T**op-$k$ **R**epresentative **S**imilar **S**ubtrajectory is named as TRSSQ, which is proposed by Wang et al [14]. Given a query trajectory $T^q$, TRSSQ problem seeks to retrieve the top-$k$ representative similar subtrajectories from a trajectory database. Here, as illustrated in the example, representative subtrajectory refers to the subtrajectory most similar to the query $T^q$ within each trajectory.

**Existing Studies.** A naive approach to solving the TRSSQ problem is to enumerate all subtrajectories of each trajectory and compute their similarity to the query. Then, it can gain representative similar subtrajectories and choose the top-$k$ results. To the best of our knowledge, Wang et al. [14] propose the most effective and so far the only approach directly targeting the TRSSQ problem. It follows the filter-validate paradigm, using a machine learning model for filtering and performing result computation during validation.

**Challenges and our Solution.** Existing frameworks for TRSSQ problem predominantly adopt the filtering-validation paradigm. However, the filtering step, despite its critical role, suffers from several limitations: 1) The filtering stage necessitates processing the entire trajectory database. 2) The filtering strategy lacks sufficient granularity, resulting in an excessively large candidate set. These limitations impose a significant computational burden and harm efficiency. Specifically, Wang et al. [14] propose the Representative Similarity Score Estimation (RSSE) model, which estimates trajectory similarity with following stages: GRU-based point-level matching, trajectory encoding and aligning, and local similarities computing. RSSE first encodes all trajectories in the database, resulting in significant computational overhead. Moreover, due to potential deviations in the similarity scores estimated by the model, it further performs top-$k$ analysis over a large candidate set to ensure the accuracy. For the TRSSQ problem, overcoming the above challenges involves mitigating the impact of large database scale on query efficiency. However, no practical solution has yet been proposed in the existing literature.

To solve this, we first observe that the data trajectories containing the top-$k$ representative similar subtrajectories for the same query exhibit high similarity, as the top-$k$ subtrajectories that are similar to the same query are likely to be similar to each other as well. Motivated by this, we propose to explore a novel indexing strategy to effectively address the TRSSQ problem. The basic idea is to build a graph index where nodes represent data trajectories, and edges and edges prioritize clustering those with top-$k$ subtrajectories of the same query. During querying, retrieval is done on the graph, avoiding full data processing during filtering and computational burden in validation.

To the best of our knowledge, there is currently no existing work that leverages graph structures for top-$k$ subtrajectory search. The most technically related work is the HNSW [15] algorithm proposed for the KNN problem. Several key issues still need to be addressed when applying graphs to TRSSQ scenarios. How should we define connections between trajectory nodes in the graph? Specifically, given an edge between $T^i$ and $T^j$, if $T^i$'s representative subtrajectory appears in the top-k query results, $T^j$ should also be likely to contribute its subtrajectory to the same top-$k$ set. Second, constructing such a graph may be inefficient. Computing the similarity between every pair of subtrajectories from $T^i$ and $T^j$ is extremely expensive, especially when dealing with large trajectory databases. Third, the graph construction and search process should guarantee convergence to the global optimum while escaping local optima. Finally, there is a subtle trade-off between query efficiency and search accuracy. To address these issues, we first introduce the DTSM(**D**ata **T**rajectory **S**imilarity **M**etric) algorithm to determine the connection relationships between trajectory nodes, ensuring that data trajectories containing the same query's top-$k$ subtrajectories are effectively clustered together in a graph. Second, by incorporating R-tree, grid filtering, and efficient pruning rules of DTSM for assessing node connectivity, the time required for offline graph index construction is substantially reduced, ensuring scalability and efficiency in handling large datasets. Third, the search process on the graph index demonstrates a certain degree of convergence. Moreover, each node in the graph index is connected to a certain percentage of randomly selected nodes as neighbors to address the issue of local optimality, thereby improving the robustness of the search process. Finally, inspired by the HNSW algorithm [15], we design a sophisticated dual-layer graph structure that significantly reduces query time and enhances performance. The upper-level graph is constructed by selecting representative trajectories from a global set of trajectories, aiming to facilitate fast retrieval and the selection of optimal search starting points for the lower-level graph. The lower-level graph is used for precise retrieval to improve accuracy. It records every node in the dataset to ensure accuracy. By controlling the number of neighboring nodes for each node in the lower-level graph, a balance is struck between query efficiency and search accuracy.

We summarize our contributions as follows:

• We propose GTRSS (**G**raph-based **T**op-$k$ **R**epresentative **S**imilar **S**ubtrajectory Query), the first pioneering framework that leverages a dual-layer graph index structure with a randomized strategy to efficiently and effectively address the top-$k$ representative similar subtrajectory query problem.

• We introduce the DTSM algorithm, a dynamic programming approach with powerful pruning rules to determine the connectivity between trajectory nodes in a graph. This approach ensures effective clustering of data trajectories containing the query's top-$k$ subtrajectories, thereby enhancing the robustness of the search process. By combining DTSM's pruning strategy with external filtering techniques, the time required to construct the offline graph index is significantly reduced, enhancing efficiency and scalability for large-scale trajectory datasets.

• Extensive experimental evaluations on real-world datasets validate the superiority of GTRSS in both efficiency and accuracy. Our approach outperforms existing state-of-the-art methods, achieving a speedup of up to 2 orders of magnitude and superior accuracy in processing large-scale trajectory data.

## II. RELATED WORK

To date, there is only one existing work [14] that directly addresses the TRSSQ problem. However, the following two related research directions are worth noting.

**Similar Subtrajectory Search.** The Similar Subtrajectory (SimSub) search problem, introduced by Wang et al. [16], aims to identify the most query-similar segment in a context trajectory. Previous work [10] proposes a two-phase filtering-validation approach to retrieve the most similar subtrajectory from large datasets. Research [9, 17, 18] optimizes the validation phase by finding the most similar subtrajectory from a given data trajectory in response to a query. For example, Wang et al. [9] introduce algorithms like ExactS, RLS, and RLS-Skip for efficient intra-trajectory search. ExactS exhaustively enumerates all candidate subtrajectories while progressively computing similarity with shared prefixes, achieving exact solutions in $O(mn^2)$ time. RLS and RLS-Skip approximate solutions via reinforcement learning to determine optimal split points, improving efficiency. Jin et al. [17] propose a conversion-matching algorithm with $O(mn)$ complexity for WED/DTW metrics by finding least-cost conversion paths. However, this method fails for LCSS/LORS metrics and requires distinct transition cost functions per metric. Despite progress, existing SimSub approaches often compromise retrieval quality in terms of subtrajectory length.

**Top-k Trajectory Similarity Search.** Top-k trajectory similarity search prioritizes efficiency and scalability. Methods broadly fall into learning and indexing categories. In learning, Aries [19] improves similarity via metric learning, though most methods focus on full trajectories rather than subtrajectories. Indexing techniques include UTgrid [20] for uncertain trajectories and RP-Trie Index [21] for local search efficiency via trie structures. Pan et al. [22] propose distributed dynamic indexing for large-scale real-time updates. While numerous methods focus on full trajectories [23–26], subtrajectory queries [27, 28] remain underexplored. To the best of our knowledge, no existing work applies graph indexing to top-$k$ subtrajectory similarity search. In contrast, the KNN field widely adopts HNSW [15], which organizes data points via a multi-level, sparse graph structure and accelerates approximate nearest neighbor queries through hierarchical search. However, applying graph indexing to the TRSSQ problem introduces challenges as mentioned.

## III. PRELIMINARIES

In this section, we provide precise definitions of key terms and concepts, followed by a description of the trajectory similarity metrics utilized in our study.

### A. Problem Definition

**Definition 1** (Trajectory). A trajectory of length $n$ is a sequence of spatial positions recorded over time, represented as $T = (p_1, p_2, \ldots, p_n)$. Each point $p_i = (t_i, lon_i, lat_i)$ typically includes a timestamp $t_i$, a longitude $lon_i$, and a latitude $lat_i$. To simplify notation, the time component $t_i$ is often omitted. A trajectory database consists of $N$ trajectories, denoted as $D = \{T^{d_1}, T^{d_2}, \ldots, T^{d_N}\}$.

**Definition 2** (Subtrajectory). For a given trajectory $T = (p_1, p_2, \ldots, p_n)$, a subtrajectory $T_{i:j}$ is a continuous portion of $T$ starting at $p_i$ and ending at $p_j$, where $1 \leq i < j \leq n$.

**Definition 3** (Representative Similar Subtrajectory). Given a data trajectory $T^d$ of length $n$ in the database $D$ and a query trajectory $T^q$ of length $m$ (where $n > m$), the representative similar subtrajectory $R^d$ is the subtrajectory of $T^d$ that has the highest similarity to $T^q$ based on a predefined similarity measure $\Theta(\cdot, \cdot)$. Formally, $R^d = T^d_{a:b}$, where $(a, b) = \underset{1 \leq a \leq b \leq n}{\arg\max} \Theta(T^q, T^d_{a:b})$ and $\Theta(T^q, R^d)$ is the representative similarity score.

**Definition 4** (Top-$k$ Representative Similar Subtrajectory Query). Given a query trajectory $T^q$, a trajectory database $D$ containing $N$ trajectories, and a similarity measure $\Theta(\cdot, \cdot)$, let $R = \{R^1, R^2, \ldots, R^N\}$ be the set of representative similar subtrajectories extracted from each data trajectory in $D$. The goal is to find a subset $R' \subseteq R$ of size $k$ such that the total similarity score is maximized:

$$R' = \underset{R' \subseteq R, |R'|=k}{\arg\max} \sum_{x \in R'} \Theta(x, T^q) \qquad (1)$$

Here, various metrics can be employed to gauge the similarity of trajectories. In this study, we select three widely recognized similarity metrics: DTW, EDR, and ERP. We denote the similarity functions as $\Theta^{DTW}$, $\Theta^{ERP}$, and $\Theta^{EDR}$.

## IV. METHOLOGY GTRSS

### A. Overview

We propose an innovative framework, GTRSS, which constructs a dual-layer graph index structure to solve the TRSSQ problem efficiently. As illustrated in Fig. 2, GTRSS includes the RNSA (Representative Node Selection and Aggregation) and CNFI (Comprehensive Node Fine-grained Integration) phases, corresponding to the construction of the upper-layer and lower-layer indices, respectively.

To construct the dual-layer graph index, we first build an R-tree on the trajectory dataset $D$ to facilitate spatial filtering. The resulting filtered dataset is denoted as $D_g$, which serves as the input for constructing the upper-layer graph index. To evaluate the similarity between trajectories, we propose a tailored correlation scoring algorithm named **DTSM** (**D**ata **T**rajectory **S**imilarity **M**etric). Based on DTSM scores, a representative node aggregation mechanism selects a portion of candidate neighbors from multiple score intervals as neighbors for each trajectory $T^{d_i} \in D_g$, forming the upper-layer graph called the **Global Aggregated Representative Index (GARI)**. GARI enables efficient retrieval of subtrajectories from the global dataset that are relatively similar to the query trajectory $T^q$.

In the CNFI phase, the entire trajectory dataset is used to build the lower-layer graph. First, for each trajectory $T^{d_i} \in D$, we select spatially closest trajectories and random trajectories as candidates. These candidates are then evaluated using the DTSM algorithm to compute their correlation scores with $T^{d_i}$. Subsequently, a comprehensive node aggregation mechanism selects a final set of neighbors from both spatially close and randomly sampled candidates to construct the lower-layer graph, referred to as the **Comprehensive Node Detail Index (CNDI)**. CNDI provides fine-grained accuracy in retrieving the top-$k$ representative similar subtrajectories.

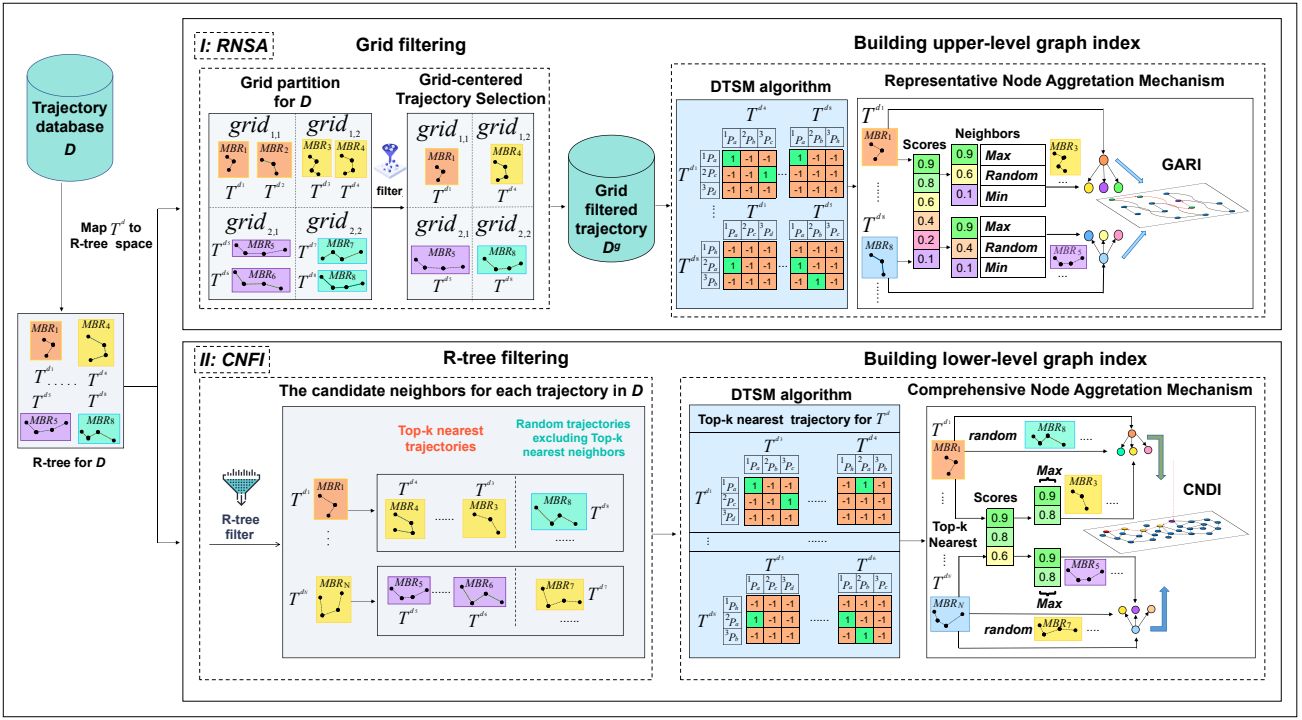

Fig 2: **The offline indexing of GTRSS is divided into two parts: RNSA and CNFI. The RNSA constructs the global aggregated representative index, GARI. The CNFI constructs the comprehensive node detail index, CNDI.**

Subsequently, online search operations are performed on both GARI and CNDI to achieve efficient and accurate retrieval of top-$k$ subtrajectories.

### B. DTSM algorithm

As mentioned, in TRSSQ, each node in dual-layer graph index of GARI and CNDI represents a trajectory from the trajectory database. During the graph construction, it is important to ensure that if a subtrajectory of $T^{d_1}$ belongs to the top-$k$ subtrajectories of $T^q$, the subtrajectories associated with the neighboring nodes of $T^{d_1}$ are also highly likely to be included in the top-$k$ results for $T^q$. To tackle this, in this subsection, we propose a novel **D**ata **T**rajectory **S**imilarity **M**etric, DTSM. Unlike traditional similarity metrics, DTSM identifies subtrajectories to calculate the similarity, which aligns with the practical needs of TRSSQ.

*1) A Metric for Subtrajectory Similarity:* We define a novel similarity evaluation function for DTSM that computes the maximum similarity score among all possible subtrajectory pairs between $T^{d_1}$ and $T^{d_2}$:

$$\phi(T^{d_1}, T^{d_2}) = \max \left\{ \Theta_s(T^{d_1}_{i:k}, T^{d_2}_{j:l}) \mid (i,k,j,l) \in \mathcal{I} \right\} \quad (2)$$

where

$$\mathcal{I} = \{(i,k,j,l) \mid 1 \leq i \leq k \leq n_1,\ 1 \leq j \leq l \leq n_2\},$$

and $\Theta_s(\cdot, \cdot)$ is a subtrajectory similarity score function that quantifies the similarity between two subtrajectories.

We refer to the pair of subtrajectories corresponding to the maximum score in Eq. (2) as the *most similar subtrajectory pair* between $T^{d_1}$ and $T^{d_2}$.

**Example 2.** *For example, as shown in fig 3(a), subtrajectories $T^{d_1}_{3:6}$ and $T^{d_2}_{1:4}$ are highly similar but may be missed by conventional methods that only consider overall similarity. By applying the proposed function $\phi(T^{d_1}, T^{d_2})$, such local similarities can be captured. A naive baseline approach would exhaustively compute the similarity scores for all possible subtrajectory pairs and return the maximum, including comparisons such as $\Theta_s(T^{d_1}_{1:1}, T^{d_2}_{1:1}), ..., \Theta_s(T^{d_1}_{1:1}, T^{d_2}_{1:8}), \Theta_s(T^{d_1}_{1:1}, T^{d_2}_{2:2}), ..., \Theta_s(T^{d_1}_{1:8}, T^{d_2}_{8:8}), \Theta_s(T^{d_1}_{2:2}, T^{d_2}_{1:1}), ...,$ up to $\Theta_s(T^{d_1}_{8:8}, T^{d_2}_{8:8})$.*

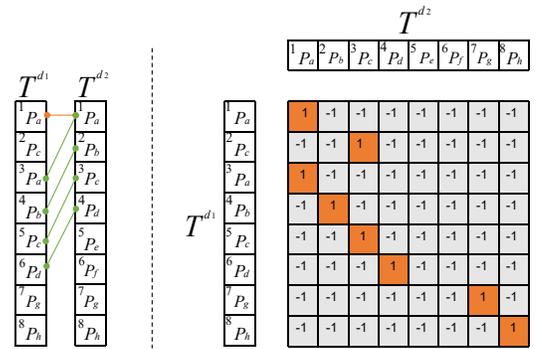

(a) Trajectory $T^{d_1}$ and $T^{d_2}$   (b) Points similarity matrix between $T^{d_1}$ and $T^{d_2}$

Fig 3: **Diagram of the Subtrajectory Matching Problem**

Clearly, evaluating $\phi(T^{d_1}, T^{d_2})$ in Equation 2 requires computing $O(n_1^2 n_2^2)$ subtrajectory pairwise similarity in the naive approach. The existing methods are unable to effectively solve these problems [29] [30]. Firstly, these algorithms [31] [32] struggle with selecting the most similar subtrajectory pair, as shown in fig 3(a), subtrajectories $T^{d_1}_{1:1} = <p_a>$ and

$T^{d_2}_{1:1} = <p_a>$, as well as $T^{d_1}_{3:6} = <p_a, p_b, p_c, p_d>$ and $T^{d_2}_{1:4} = <p_a, p_b, p_c, p_d>$, are treated as equally similar, with algorithms assuming identical similarity scores without considering the number of similar point pairs. In addition, trajectory similarity methods such as DTW [4] and LCSS [33] focus on aligning similar points, ignoring the dissimilar ones. This may lead to neglect local similarities, focusing instead on overall trajectory similarity. $T^{d_1}_{3:7}$ and $T^{d_2}_{1:5}$ are more suitable as the most similar subtrajectory pair than $T^{d_1}_{3:8}$ and $T^{d_2}_{1:6}$, as the former contains fewer dissimilar points.

*2) Subtrajectory Similarity evaluation:* To address above issue, we propose a concept of the Point Pair Matching Rules.

**Definition 6** (Point Pair Matching Rules). Given two points $p^{d_1}_i$ and $p^{d_2}_j$, the rules for point pair matching are defined as follows:

• **Corresponding Points**: If distance $d(p^{d_1}_i, p^{d_2}_j) \leq \alpha$, where $\alpha$ is a predefined threshold, then these two points are considered corresponding points and are matched. Once matched, these points are no longer available for further pairing in subsequent steps.

• **Non-Corresponding Points**: If $p^{d_1}_i$ and $p^{d_2}_j$ are not corresponding points, a secondary comparison is performed: $p^{d_1}_i$ is compared with $p^{d_2}_{j+1}$, and $p^{d_2}_j$ is compared with $p^{d_1}_{i+1}$. If no matching is found in this secondary comparison, the pair is considered non-corresponding points and will be excluded from subsequent matching attempts.

• **Counting Point Pairs**: $|C(T^{d_1}_{i:k}, T^{d_2}_{j:l})|$ and $|D(T^{d_1}_{i:k}, T^{d_2}_{j:l})|$ denote the respective numbers of corresponding and non-corresponding point pairs between the subtrajectories $T^{d_1}_{i:k}$ and $T^{d_2}_{j:l}$, respectively.

Based on the above matching rules, the similarity score between two subtrajectories $T^{d_1}_{i:k}$ and $T^{d_2}_{j:l}$ is proportional to:

$$\Theta_s(T^{d_1}_{i:k}, T^{d_2}_{j:l}) \propto |C(T^{d_1}_{i:k}, T^{d_2}_{j:l})| - |D(T^{d_1}_{i:k}, T^{d_2}_{j:l})| \quad (3)$$

This similarity score offers a robust metric that avoids overestimating similarity while penalizing mismatched points. In addition, this rule can effectively address the issue illustrated in Fig. 3(a). When the number of similar point pairs between $T^{d_1}_{3:6}$ and $T^{d_2}_{1:4}$ is greater than that between $T^{d_1}_{1:1}$ and $T^{d_2}_{1:1}$, we prioritize selecting $T^{d_1}_{3:6}$ and $T^{d_2}_{1:4}$ as a candidate for the most similar subtrajectory pair. Similarly, we select the subtrajectories $T^{d_1}_{3:7}$ and $T^{d_2}_{1:5}$, as they contain fewer dissimilar point pairs compared to $T^{d_1}_{3:8}$ and $T^{d_2}_{1:6}$.

*3) Efficient Computation of the Most Similar Subtrajectory Pair:* According to the Point Pair Matching Rules defined in Definition 6, an appropriate similarity measure is provided for evaluating subtrajectory similarity. However, the issue of high computational complexity still remains. As mentioned above, $\phi(T^{d_1}, T^{d_2})$ defined in Equation 2 requires evaluating $n_1^2 \cdot n_2^2$ subtrajectory pairs. The computational complexity of comparing each pair is at least $O(n_1 n_2)$. Consequently, the overall time complexity of this process, when using exhaustive approaches such as Exact, reaches at least $O(n_1^3 n_2^3)$. Even with more efficient algorithms such as PSS [9] or CMA [17], the computational cost remains prohibitively high, making tasks such as graph indexing impractical due to the significant processing time involved.

To efficiently identify the most similar subtrajectory pair between two trajectories $T^{d_1}$ and $T^{d_2}$, we design a dynamic programming (DP) algorithm based on a similarity matrix and a recursive scoring function. This approach avoids redundant computations and supports effective pruning, significantly reducing computational cost.

We first define a binary matrix $A \in \{-1, 1\}^{n_1 \times n_2}$, where each element is determined as follows:

$$A_{i,j} = \begin{cases} 1 & \text{if } d(p^{d_1}_i, p^{d_2}_j) \leq \alpha \\ -1 & \text{otherwise} \end{cases} \quad (4)$$

Each entry $A_{i,j}$ indicates whether the point pair $(p^{d_1}_i, p^{d_2}_j)$ is similar under the distance threshold $\alpha$, as illustrated in Fig. 3(b).

The similarity between two subtrajectories is defined recursively using Equation 5, with pairwise similarity assessed by the following scoring function:

$$\Theta_s(T^{d_1}_{i:k}, T^{d_2}_{j:l}) = \begin{cases} 0 & \text{if } d(p^{d_1}_i, p^{d_2}_j) > \alpha \\ subcost(p^{d_1}_i, p^{d_2}_j) \\ + subcost(p^{d_1}_{i+1}, p^{d_2}_{j+1}) & \text{otherwise} \\ + R(T^{d_1}_{i+1:k}, T^{d_2}_{j+1:l}) \end{cases} \quad (5)$$

where

$$subcost(p^{d_1}_i, p^{d_2}_j) = \begin{cases} 2 & \text{if } d(p^{d_1}_i, p^{d_2}_j) \leq \alpha \\ -2 & \text{otherwise} \end{cases}$$

The recursive term $R(T^{d_1}_{i+1:k}, T^{d_2}_{j+1:l})$ selects the optimal path under several matching conditions $C1$ through $C6$, shown in Equation 6. Each condition reflects a different alignment scenario, illustrated in Fig. 4.

$$R(T^{d_1}_{i+1:k}, T^{d_2}_{j+1:l}) =$$
$$\begin{cases} costA(P_{2,2}) + R(T^{d_1}_{i+2:k}, T^{d_2}_{j+2:l}) & \text{if } C1 \\ costB(P_{1,2}) + R(T^{d_1}_{i+1:k}, T^{d_2}_{j+2:l}) & \text{if } C2 \\ costC(P_{2,1}) + R(T^{d_1}_{i+2:k}, T^{d_2}_{j+1:l}) & \text{if } C3 \\ \max \begin{cases} costB(P_{1,2}) + R(T^{d_1}_{i+1:k}, T^{d_2}_{j+2:l}) \\ costC(P_{2,1}) + R(T^{d_1}_{i+2:k}, T^{d_2}_{j+1:l}) \end{cases} & \text{if } C4 \\ costD(P_{2,2}) + R(T^{d_1}_{i+2:k}, T^{d_2}_{j+2:l}) & \text{if } C5 \\ \max \begin{cases} costE(P_{1,2}) + R(T^{d_1}_{i+1:k}, T^{d_2}_{j+2:l}) \\ costF(P_{2,1}) + R(T^{d_1}_{i+2:k}, T^{d_2}_{j+1:l}) \end{cases} & \text{if } C6 \end{cases} \quad (6)$$

where
$$\begin{cases} C1: & d(P_{1,1}) \leq \alpha \\ C2: & d(P_{1,1}) > \alpha \land d(P_{1,2}) \leq \alpha \land d(P_{2,1}) > \alpha \\ C3: & d(P_{1,1}) > \alpha \land d(P_{1,2}) > \alpha \land d(P_{2,1}) \leq \alpha \\ C4: & d(P_{1,1}) > \alpha \land d(P_{1,2}) \leq \alpha \land d(P_{2,1}) \leq \alpha \\ C5: & d(P_{1,1}) > \alpha \land d(P_{1,2}) > \alpha \land d(P_{2,1}) > \alpha \land d(P_{2,2}) \leq \alpha \\ C6: & d(P_{1,1}) > \alpha \land d(P_{1,2}) > \alpha \land d(P_{2,1}) > \alpha \land d(P_{2,2}) > \alpha \end{cases}$$

To simplify notation, we denote matched point pairs as $P_{u,v} = (p^{d_1}_{i+u}, p^{d_2}_{j+v})$, where $u, v \in \{1, 2\}$.

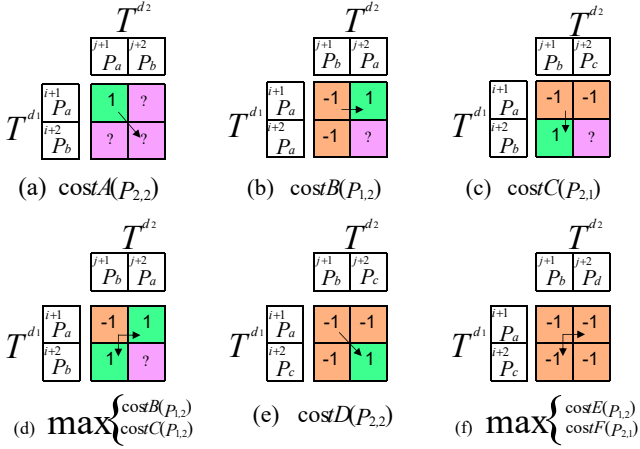

Fig 4: **DTSM recursive strategy under different matching conditions**

As shown in Fig 4, each condition corresponds to a specific recurrence case:

**Condition C1.** (Fig. 4(a)): A similar point pair is found at $(p_{i+1}^{d_1}, p_{j+1}^{d_2})$, which contributes 1 to the similarity score of the current subtrajectory pair. Since this pair is already matched, it will not participate in subsequent similarity calculations. The algorithm then proceeds to evaluate the next point pair $(p_{i+2}^{d_1}, p_{j+2}^{d_2})$, and its contribution is computed as follows:

$$costA(P_{2,2}) = subcost(P_{2,2}) \quad (7)$$

The remaining similarity is calculated recursively using $R(T_{i+2:k}^{d_1}, T_{j+2:l}^{d_2})$.

**Condition C2.** (Fig. 4(b)): The point pair $(p_{i+1}^{d_1}, p_{j+1}^{d_2})$ is dissimilar and contributes $-1$, but the pair $(p_{i+1}^{d_1}, p_{j+2}^{d_2})$ is similar. To account for this, $p_{i+1}^{d_1}$ will recalculate the contribution value of the subtrajectory similarity together with $p_{j+2}^{d_2}$, and the previous contribution value of $p_{i+1}^{d_1}$ is cancelled, because it is $-1$ before, using $\Delta()$ to add 1 now means canceling the previous contribution. And then receive new contributions:

$$costB(P_{1,2}) = subcost(P_{1,2}) + \Delta(p_{i+1}^{d_1}) \quad (8)$$

where $\Delta(p_{i+1}^{d_1}) = 1$, then use $R(T_{i+1:k}^{d_1}, T_{j+2:l}^{d_2})$ to calculate the remaining point pairs.

**Condition C3.** (Fig. 4(c)): This case is symmetric to Condition C2. The pair $(p_{i+1}^{d_1}, p_{j+1}^{d_2})$ is dissimilar, but $(p_{i+2}^{d_1}, p_{j+1}^{d_2})$ is similar. The negative contribution of $p_{j+1}^{d_2}$ is canceled and replaced with a positive one:

$$costC(P_{2,1}) = subcost(P_{2,1}) + \Delta(p_{j+1}^{d_2}) \quad (9)$$

where $\Delta(p_{j+1}^{d_2}) = 1$. The recursion proceeds with $R(T_{i+2:k}^{d_1}, T_{j+1:l}^{d_2})$.

**Condition C4.** (Fig. 4(d)): Both $(p_{i+1}^{d_1}, p_{j+2}^{d_2})$ and $(p_{i+2}^{d_1}, p_{j+1}^{d_2})$ are similar, offering two valid options for adjustment. The algorithm chooses the one yielding the higher total similarity:

$$\max \begin{cases} costB(P_{1,2}) + R(T_{i+1:k}^{d_1}, T_{j+2:l}^{d_2}), \\ costC(P_{2,1}) + R(T_{i+2:k}^{d_1}, T_{j+1:l}^{d_2}) \end{cases} \quad (10)$$

**Condition C5.** (Fig. 4(e)): All three adjacent point pairs are dissimilar, but $(p_{i+2}^{d_1}, p_{j+2}^{d_2})$ is similar. In this case, the contribution is simply:

$$costD(P_{2,2}) = subcost(P_{2,2}) \quad (11)$$

with the recursion proceeding to $R(T_{i+2:k}^{d_1}, T_{j+2:l}^{d_2})$.

**Condition C6.** (Fig. 4(f)): None of the considered point pairs are similar. To find an improved alignment, the algorithm explores two alternative paths and selects the one with the higher score:

$$\max \begin{cases} costE(P_{1,2}) + R(T_{i+1:k}^{d_1}, T_{j+2:l}^{d_2}), \\ costF(P_{2,1}) + R(T_{i+2:k}^{d_1}, T_{j+1:l}^{d_2}) \end{cases}$$

where

$$costE(P_{1,2}) = subcost(P_{1,2}) + \Delta(p_{i+1}^{d_1}) \quad (12)$$

$$costF(P_{2,1}) = subcost(P_{2,1}) + \Delta(p_{j+2}^{d_2}) \quad (13)$$

and $\Delta(p_{i+1}^{d_1}) = \Delta(p_{j+2}^{d_2}) = 1$.

**Pruning Rule.** When traversing the matrix $A$, if a starting point pair $(p_i^{d_1}, p_j^{d_2})$ is dissimilar (i.e., $d(p_i^{d_1}, p_j^{d_2}) > \alpha$), the algorithm skips all subtrajectory comparisons that begin from this pair. We now prove that pruning these subtrajectories does not miss any optimal subtrajectory pair.

**Theorem.** Let $\Theta_s(T_{i:k}^{d_1}, T_{j:l}^{d_2})$ be defined as in Equation 5. If $d(p_i^{d_1}, p_j^{d_2}) > \alpha$, then:

$$\Theta_s(T_{i:k}^{d_1}, T_{j:l}^{d_2}) \leq \Theta_s(T_{i+1:k}^{d_1}, T_{j+1:l}^{d_2})$$

*Proof.* We use induction on the subtrajectory length $L = \min(k - i + 1, l - j + 1)$.

**Base case:** $L = 1$. Then $\Theta_s(T_{i:i}^{d_1}, T_{j:j}^{d_2}) = 0$ because $d(p_i^{d_1}, p_j^{d_2}) > \alpha$, and $\Theta_s(T_{i+1:i+1}^{d_1}, T_{j+1:j+1}^{d_2}) \geq 0$, so the inequality holds.

**Inductive step:** Assume the claim holds for subtrajectories of length up to $L$. Consider a subtrajectory of length $L + 1$. Again, since $d(p_i^{d_1}, p_j^{d_2}) > \alpha$, the similarity is reset to 0 by Equation 5:

$$\Theta_s(T_{i:k}^{d_1}, T_{j:l}^{d_2}) = 0$$

Meanwhile, the similarity from $T_{i+1:k}^{d_1}, T_{j+1:l}^{d_2}$ may be nonzero, therefore the inequality holds. ∎

This result ensures that no high-scoring subtrajectory pair will be missed by pruning at dissimilar starting points. Hence, the pruning strategy is both valid and optimal under the scoring function $\Theta_s$.

**Computational Complexity.** The naive method requires evaluating $O(n_1^2 n_2^2)$ subtrajectory pairs, with each requiring $O(n_1 n_2)$ time. The total complexity becomes $O(n_1^3 n_2^3)$. With dynamic programming and memoization, we reduce this to $O(n_1^2 n_2^2)$, and the pruning strategy further brings the expected cost closer to $O(n_1 n_2)$ in practice.

**Algorithm 1:** DTSM Algorithm

**Input** : Two trajectories $T^{d_1}$ and $T^{d_2}$ of lengths $n_1$ and $n_2$; threshold $\alpha$
**Output:** The maximum subtrajectory similarity score $S_{\max}$

1 Initialize matrix $A \in \{-1,1\}^{n_1 \times n_2}$ using Equation 4
2 $S_{\max} \leftarrow 0$
3 **for** $i = 1$ **to** $n_1$ **do**    // Scan all starting point pairs
4    **for** $j = 1$ **to** $n_2$ **do**
5       **if** $A_{i,j} = -1$ **then**  // Prune dissimilar start points
6          **continue**
7       **end**
8       $S \leftarrow 0, \quad S_{\max}^{\text{sub}} \leftarrow 0$
9       **for** $k = i$ **to** $n_1$ **do**
10          **for** $l = j$ **to** $n_2$ **do**
11             $S \leftarrow \Theta_s(T^{d_1}_{i:k}, T^{d_2}_{j:l})$
12             **if** $S > S_{\max}^{\text{sub}}$ **then**
13                $S_{\max}^{\text{sub}} \leftarrow S$
14             **end**
15             **if** $S \leq 0$ **then**
16                **break**
17             **end**
18          **end**
19       **end**
20       **if** $S_{\max}^{\text{sub}} > S_{\max}$ **then**
21          $S_{\max} \leftarrow S_{\max}^{\text{sub}}$
22       **end**
23    **end**
24 **end**
25 **return** $S_{\max}$

*4) Summary and Insights on the DTSM Algorithm:* We summarize the overall DTSM computation in Algorithm 1. The DTSM algorithm traverses the similarity matrix $A$, and for each similar starting point pair $(p_i^{d_1}, p_j^{d_2})$, it computes the similarity score of all possible subtrajectory pairs beginning at that point. To avoid redundant computation, we adopt a dynamic programming strategy: each similarity score is incrementally updated based on the results of previously computed subproblems, following the recursive formulation in Equation 5. We also apply early termination: if the similarity score drops to zero or below, further extension of the current subtrajectory is skipped, since it cannot produce a higher score. This mechanism, along with pruning based on dissimilar starting points, significantly reduces unnecessary calculations while ensuring optimal subtrajectory similarity is found.

### C. Offline indexing

The offline indexing of GTRSS plays a crucial role in supporting the Top-$k$ Representative Similar Subtrajectory Query. As illustrated in Fig. 2, the indexing framework adopts a dual-layer graph structure, consisting of two key components: the Representative Node Set Aggregation (RNSA) and the Comprehensive Node Feature Indexing (CNFI). The RNSA component is responsible for constructing the Global Aggregated Representative Index (GARI), which captures high-level structural summaries. In contrast, CNFI focuses on generating the Comprehensive Node Detail Index (CNDI), which preserves fine-grained information at the node level. In the following, we provide a detailed analysis of the construction processes of both RNSA and CNFI.

*1) RNSA: Representative Node Set Aggregation:* To ensure efficient search in the CNFI graph, the selection of an effective starting node is crucial. An inappropriate starting trajectory may require numerous transitions before converging to a highly similar subtrajectory, which increases computational overhead and risks falling into local optima. To address this, the RNSA component constructs GARI, a sparse but globally representative graph of trajectories that enables fast location of promising candidates.

The core idea of RNSA is to first divide the trajectory dataset spatially and select representative trajectories from each spatial region. Then, it connects each representative node to a set of neighbors that cover local similarity, global exploration, and random transitions. This design facilitates both accurate and broad search space coverage.

The RNSA process consists of two main stages: *grid filtering* and *upper-level index building*, as illustrated in Fig. 2. In the grid filtering stage, we first determine the spatial extent of the trajectory dataset by computing the minimum and maximum coordinates along both spatial dimensions. The entire space is then partitioned into $M \times M$ uniform grid cells. Within each cell, a centered trajectory is selected through R-tree filtering. This ensures that the resulting set of representatives is both spatially diverse and compact, reducing redundancy while preserving overall spatial coverage.

In the upper-level index building stage, we first compute pairwise similarities among the selected representative trajectories using the proposed DTSM algorithm. This similarity information enables the construction of a robust neighbor structure. Specifically, for each representative trajectory, we design a dedicated node aggregation mechanism that identifies three categories of neighboring trajectories: (i) the most similar trajectory, which ensures high local fidelity within the index; (ii) a randomly selected trajectory, which enhances exploration capability and helps prevent the search process from getting trapped in local optima; and (iii) the least similar trajectory, which promotes broader global coverage across the trajectory space. This hybrid neighbor selection strategy effectively balances local consistency and global diversity. As a result, it forms the GARI, which serves as the upper layer of the offline indexing framework and provides high-quality candidate nodes for subsequent fine-grained similarity search.

The RNSA algorithm is shown in Algorithm 2. It begins by computing the spatial boundaries of the dataset (Lines 1–6) and partitions the space into uniform grids (Lines 7–10). It then applies R-tree filtering to extract a representative trajectory from each grid cell (Lines 11–16). After collecting

**Algorithm 2: RNSA Algorithm**

**Input** : Trajectory dataset $D = \{T^{d_1}, T^{d_2}, \ldots, T^{d_N}\}$.
**Output:** GARI $G^t$.

1. Initialize the spatial coordinates of the data trajectory set with $x_{min} \leftarrow \infty$, $y_{min} \leftarrow \infty$, $x_{max} \leftarrow 0$, $y_{max} \leftarrow 0$
2. **for** each $T^{d_i} \in D$ **do**
3.    **for** each $p_j \in T^{d_i}$ **do**
4.       Update $\{x_{min}, y_{min}, x_{max}, y_{max}\}$ based on $(x_j, y_j)$ of $p_j$
5.    **end**
6. **end**
7. Initialize the number of single-sided grids $M$ in the space
8. Calculate the width and height of a single grid
9. $W_{Grid} \leftarrow (x_{max} - x_{min})/M$
10. $H_{Grid} \leftarrow (y_{max} - y_{min})/M$
11. Initialize an empty set or matrix for $D_g = \emptyset$;
12. **for** $i = 1, \ldots, M$ **do**
13.    **for** $j = 1, \ldots, M$ **do**
14.       Extract $T^{top_k}$ from grid $G_{i,j}$ and store it in $D_g$;
15.    **end**
16. **end**
17. Obtain $D_g = \{T^{top_1}, T^{top_2}, \ldots, T^{top_n}\}$;
18. **for** each $T^{top_i} \in D_g$ **do**
19.    **for** each $T^{top_j} \in D_g$ **do**
20.       Calculate the similarity score $\phi(T^{top_i}, T^{top_j})$
21.    **end**
22. **end**
23. **for** each $T^{top_i} \in D_g$ **do**
24.    Sort and partition $D_g$ into $D_A$, $D_B$, and $D_C$ based on similarity scores with $T^{top_i}$
25.    Select $D_{A'}$ that is closest to $T^{top_i}$ from $D_A$
26.    Randomly select $D_{B'}$ from $D_B$
27.    Select $D_{C'}$ that is farthest to $T^{top_i}$ from $D_C$
28.    Use $D_{A'}$, $D_{B'}$, $D_{C'}$ as neighbor nodes of $T^{top_i}$
29. **end**
30. Building GARI $G^t$ completed

---

**Algorithm 3: CNFI Algorithm**

**Input** : Trajectory dataset $D = \{T^{d_1}, T^{d_2}, \ldots, T^{d_N}\}$.
**Output:** CNDI $G^b$.

1. **for** each $T^{d_i} \in D$ **do**
2.    Compute $MBR$ of $T^{d_i}$ and map to R-tree space
3. **end**
4. **for** each $T^{d_i} \in D$ **do**
5.    Retrieve $D^n$ close to $T^{d_i}$ from $D$
6.    Randomly obtain $D^r$ from $D$ and not in $D^n$
7.    **for** each $T^{n_i} \in D^n$ **do**
8.       Calculate the similarity score $\phi(T^{d_i}, T^{n_i})$
9.    **end**
10.    **for** each $T^{r_i} \in D^r$ **do**
11.       Calculate the similarity score $\phi(T^{d_i}, T^{r_i})$
12.    **end**
13.    Sort $D^n$ and $D^r$ based on their similarity to $T^{d_i}$
14.    Select a specified proportion of $D^{n'}$ and $D^{r'}$ from $D^n$ and $D^r$
15.    $D^{n'}$ and $D^{r'}$ form the neighbor node of $T^{d_i}$
16. **end**
17. Building CNDI $G^b$ completed

---

representatives into $D_g$, the algorithm computes pairwise similarity scores using the DTSM method (Lines 17–22). Each representative node is then connected to a triad of neighbors that are highly similar, randomly chosen, and dissimilar, respectively (Lines 23–29). This strategy ensures that each node in GARI has access to both local and global search directions, which is crucial for overcoming local optima during subtrajectory similarity query processing.

*2) CNFI: Comprehensive Node Filtering and Indexing:* With the global representative graph GARI constructed in the RNSA phase, we can efficiently locate trajectories that are similar to the query. Building upon this, the CNFI stage constructs the CNDI, a lower-layer graph that connects each trajectory with a selected set of neighbors to enable fine-grained and efficient subtrajectory similarity retrieval.

The CNFI process consists of two main components: (1) Neighbor Candidate Filtering, which identifies a set of spatially close trajectories for each data trajectory using R-tree and random sampling, and (2) lower-level graph indexing, which selects final neighbor nodes based on similarity ranking to balance precision and exploration.

The full CNFI construction process is illustrated in Algorithm 3. In neighbor candidate filtering, for each trajectory $T^{d_i}$, an R-tree filters data trajectories to extract spatially relevant candidates, reducing the candidate set to top-$k$ neighbors $D^n$ while maintaining precision. Random neighbors from non-top-$k$ trajectories $D^r$ are also introduced to diversify the search space and prevent getting trapped in local optima. In lower-level graph indexing, the DTSM algorithm computes similarity scores between $T^{d_i}$ and each candidate in $D^n \cup D^r$. It then selects the top $\delta \cdot \xi$ trajectories from $D^n$ (high similarity) and $(1 - \delta) \cdot \xi$ trajectories from $D^r$ (random) trajectories as neighbors. This hybrid strategy ensures both search efficiency and robustness. $\delta$ tunes the ratio of similarity-guided to random neighbors, balancing between precision and diversity.

*D. Online search*

**Overview:** The online search phase consists of two parts: searching the GARI and searching the CNDI, as shown in Fig 5. The process begins by efficiently identifying the subtrajectory node in GARI that is most similar to the query trajectory. This node then serves as the entry point for a fine-grained search on CNDI, which locates the top-$k$ representative similar subtrajectories. This hierarchical retrieval strategy achieves a balance between computational efficiency, provided by global-level filtering, and retrieval accuracy, ensured through local-level refinement.

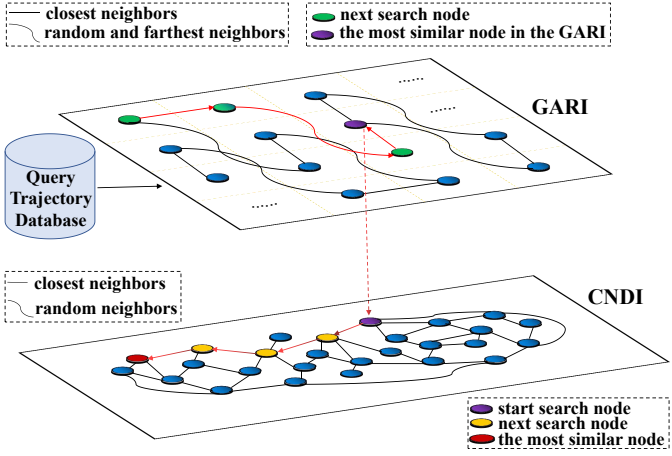

Fig 5: **Query processing during the online search phase**

**Search on GARI:** As shown in Algorithm 4, the search begins by randomly selecting a node $T^{\triangledown}$ from GARI as the initial search point (Line 3). The representative similarity score between the query trajectory $T^{q_i}$ and this node is computed. Then, the node's neighbors $C_N^{\diamondsuit}$ in the GARI graph are evaluated using similarity metrics such as Exacts, CMA [17], POS, or PSS [9]. The neighbor node with the highest similarity score exceeding that of the current node is chosen as the next node to visit. This process continues iteratively (Lines 4–11) until no neighbor yields a higher similarity score, indicating that a local maximum (denoted $T^{\blacktriangledown}$) has been reached. This node is deemed the most globally similar representative trajectory and is used as the starting point for the next stage.

**Search on CNDI:** The CNDI search starts from the node $T^{\triangle}$ obtained from GARI (Line 15). Neighbor nodes $C_N^{\boxplus}$ of the current node are retrieved and scored (Lines 16–18). Each neighbor in $C_N^{\boxplus}$ is composed of both high-similarity trajectories and randomly selected ones, allowing for a balance between exploitation and exploration. The node with the highest score is selected as the next to visit, and this process continues until all neighbors yield no higher similarity score than the current one (Line 19-23). The final node $T^{\blacktriangle}$ is considered the trajectory containing the subtrajectory that most similar to $T^{q_i}$ in the fine-grained CNDI graph. Finally, the top-$k$ representative similar subtrajectories are selected from among the candidate trajectories evaluated during the search process (Lines 26–28). This hybrid search strategy avoids local optima while ensuring efficient retrieval through a combination of hierarchical indexing and local exploration.

*E. Complexity Analysis*

**Space Complexity.** The offline indexing framework consists of two components: the GARI and the CNDI. Let $N$ denote the total number of trajectories in the dataset, and $M^2$ ($M \times M$) denote the number of representative trajectories selected during the grid partition process in RNSA. Each representative trajectory in GARI maintains a constant number of neighbors, so the space complexity of GARI is $O(m)$. For CNDI, each trajectory $T_i^d \in D$ maintains a set of $\xi$ neighbor links,

---

**Algorithm 4:** Online Search Algorithm

**Input:** Query set $Q = \{T^{q_1}, \ldots, T^{q_n}\}$, GARI $G^t$, CNDI $G^b$.
**Output:** Top-$k$ representative similar subtrajectories for each $T^{q_i}$.

1 **for** *each query trajectory $T^{q_i} \in Q$* **do**
2   // Search GARI to locate global representative
3   Randomly initialize the visited node $T^{\triangledown}$ in $G^t$
4   Compute similarity score $S_q \leftarrow Score(T^{q_i}, T^{\triangledown})$
5   Obtain $C_N^{\diamondsuit}$ (neighbors of $T^{\triangledown}$)
6   Compute $C_{score}^{\diamondsuit}$ (scores of $C_N^{\diamondsuit}$ with $T^{q_i}$)
7   **while** $\exists S_q^* \in C_{score}^{\diamondsuit} : S_q^* > S_q$ **do**
8     Select next visited node $T^{\triangledown}$ (with highest score)
9     $S_q \leftarrow S_q^{\star}$ (score of $T^{\triangledown}$ with $T^{q_i}$)
10     Obtain $C_N^{\diamondsuit}$ (neighbors of $T^{\triangledown}$)
11     Compute $C_{score}^{\diamondsuit}$ (scores of $C_N^{\diamondsuit}$ with $T^{q_i}$)
12   **end**
13   Find $T^{\blacktriangledown}$ in $G^t$ that is most similar trajectory to $T^{q_i}$
14   // Search CNDI from representative node
15   $T^{\triangle} \leftarrow T^{\blacktriangledown}$ ($T^{\triangle}$ as the starting node for CNDI)
16   Compute and record $S_q \leftarrow Score(T^{q_i}, T^{\triangle})$
17   Obtain $C_N^{\boxplus}$ (neighbors of $T^{\triangle}$)
18   Compute and record $C_{score}^{\boxplus}$
19   **while** $\exists S_q^* \in C_{score}^{\boxplus} : S_q^* > S_q$ **do**
20     Select next visited node $T^{\triangle}$ (with highest score)
21     $S_q \leftarrow S_q^{\star}$ (score of $T^{\triangle}$ with $T^{q_i}$)
22     Obtain $C_N^{\boxplus}$ (neighbors of $T^{\triangle}$)
23     Compute and record $C_{score}^{\boxplus}$
24   **end**
25   Find $T^{\blacktriangle}$ in $G^b$ that is most similar trajectory to $T^{q_i}$
26   Obtain $C_N^{\boxplus}$ (neighbors of $T^{\blacktriangle}$)
27   Obtain and record $C_{score}^{\boxplus}$
28   Return the top-$k$ $\{T_{i:k}^1, T_{j:l}^2, \ldots, T_{s:e}^k\}$ of $T^{q_i}$
29 **end**

---

consisting of both similar and randomly selected trajectories. Thus, the space required for CNDI is $O(N \cdot \xi)$. Therefore, the overall space complexity of the offline index is:

$$O(M^2 + N \cdot \xi)$$

where $M^2 \ll N$, and $\xi$ is a small constant relative to $N$.

**Time Complexity.**

*Offline Construction.* The grid-based partitioning in RNSA involves iterating over all trajectories and assigning them to spatial grids, which requires $O(N \cdot L)$ time, where $L$ is the average length of a trajectory. For each grid cell, one representative trajectory is selected, and pairwise similarity scores among the $M^2$ representatives are computed using the DTSM algorithm. Assuming DTSM has time complexity $O(L^2)$ per pair, the total cost for GARI construction is $O(M^4 \cdot L^2)$. For CNDI, each trajectory computes similarity with a subset of $\xi$ candidate neighbors, leading to a complexity of $O(N \cdot \xi \cdot L^2)$.

Thus, the overall time complexity of offline indexing is:

$$O(N \cdot L + M^4 \cdot L^2 + N \cdot \xi \cdot L^2)$$

## V. EXPRIMENTAL EVALUATION

### A. Dataset Description

We use real trajectory datasets from Xi'an and Chengdu, sourced from the DiDi Chuxing GAIA Open Dataset [34]. The datasets include 3.1 million trajectories from Xi'an in October 2016 and 5.8 million trajectories from Chengdu in November 2016. Following previous works [35] [36], we remove erroneous trajectories, ensuring that data trajectories are longer than query trajectories. Query trajectories are set between 30 and 90, and data trajectories are set between 90 and 300. We randomly select 30,000 trajectories as data trajectories and 2,000 as query trajectories from each dataset.

### B. Experimental Setup

**Baselines:** To evaluate the efficiency and effectiveness of the GTRSS framework in addressing the top-$k$ representative similar subtrajectory query problem, we use existing filtering methods for trajectory similarity as baselines.

1) OSF: This is a filtering method [10] for finding similar subtrajectories in road networks. It decreases the candidate trajectory set through minimal candidate computation and index lookup. We adapt it to Euclidean space.

2) Grid: This method divides the spatial domain into equal-sized cells. Data trajectories are assigned to these cells based on their paths. Given a query, only trajectories passing through certain cells are considered, filtering out others.

3) LBF(RSSE model): It is the state-of-the-art method for representative similar subtrajectory query, which is proposed by Wang et al [14]. It uses predicted representative similarity scores to select candidates. To reduce unnecessary predictions, we first apply the OSF filter to eliminate irrelevant trajectories.

**Metrics:** To evaluate the performance of GTRSS, we use the following metrics. (1) We use the Relative Rank (RR) metric to evaluate the quality of top-$k$ subtrajectories. For a given query, each method generates its top-$k$ subtrajectories, which are then ranked. The RR value represents the normalized average rank of these subtrajectories. Lower RR values indicate better performance. (2) HR-10 measures the correspondence between the top-10 predictions and the actual top-10 scores, reflecting the model's accuracy in identifying similar data points. Similarly, there are metrics such as HR-20, HR-30, HR-40, and HR-50, which range from 0 to 1, with values approaching 1 indicating superior performance.

**Implementation Details:** The data processing is implemented in Python 3.9 and bound to C++ code through pybind11. The methods are implemented in C++14. The experiments are conducted on a Linux server equipped with 112 CPU cores, 8GB swap, 1007GB RAM.

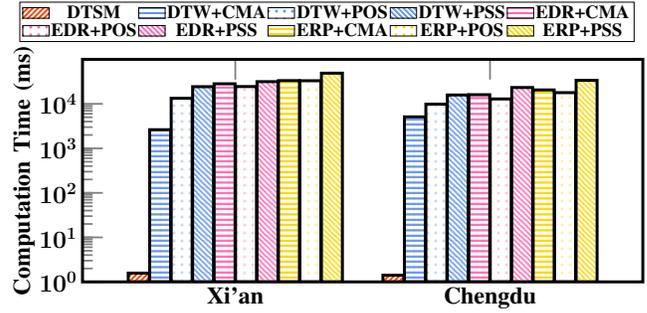

Fig 6: **Computation time for similarity measure of 10 trajectory pairs on Xi'an and Chengdu Datasets**

### C. Indexing Efficiency

The first experiment measures graph index construction time, dominated by trajectory similarity computation for neighbor determination. A comparison of similarity computation time for 10 trajectory pairs reveals: To avoid missing potential similar subtrajectories, we compute subsegment similarities of each trajectory pair, ensuring index quality.

Fig. 6 shows that DTSM computes 10 trajectory pairs in about 1ms, over 10 times faster than alternatives. Scalability tests in Section V-G also demonstrate DTSM constructs indexes for 0.3M trajectories in 12 hours and 0.8M in 30 hours.

### D. Effectiveness and Efficiency Evaluation

Fig. 7 compares the effectiveness and efficiency of our proposed GTRSS with several other approaches across different datasets. Among all evaluated methods, GTRSS+CMA achieves the best overall performance. Query are sampled from distinct length ranges: [30–45), [45–60), [60–75), and [75–90), with two trajectories per range. All methods process these queries to guarantee consistent evaluation. Experimental results show that Grid methods require approximately $10^4$ seconds ($\approx$2.8 hours) per query, while LBF and OSF methods take $10^3$ seconds ($\approx$16.7 minutes). In contrast, GTRSS+ExactS spends $10^2 \sim 10^3$ seconds (1.7–16.7 minutes), and GTRSS+CMA, GTRSS+PSS, and GTRSS+POS complete queries in $\leq 10^1$ seconds ($\leq$0.17 minutes). Notably, GTRSS+CMA achieves response times $\approx$ 2 seconds for all eight queries, outperforming the fastest prior method by over 100 times. Further experiments are discussed below.

In addition to the promising performance of the RR metric, the top-$k$ retrieval effectiveness of GTRSS is evaluated by comparing it with existing learning-based similarity estimation methodologies. We use the following methods as baselines:

1) T3S [35] uses a self-attentive-based network to process the trajectory of the Grid representation and an LSTM to encode a sequence of coordinates containing spatial information.

2) TMN [37] uses an attention mechanism to capture the connection between points in two trajectories.

3) RSSE [14] is the state-of-the-art method for representative similar subtrajectory query. It uses a deep learning model to approximate subtrajectory similarity scores and reduce the candidate set.

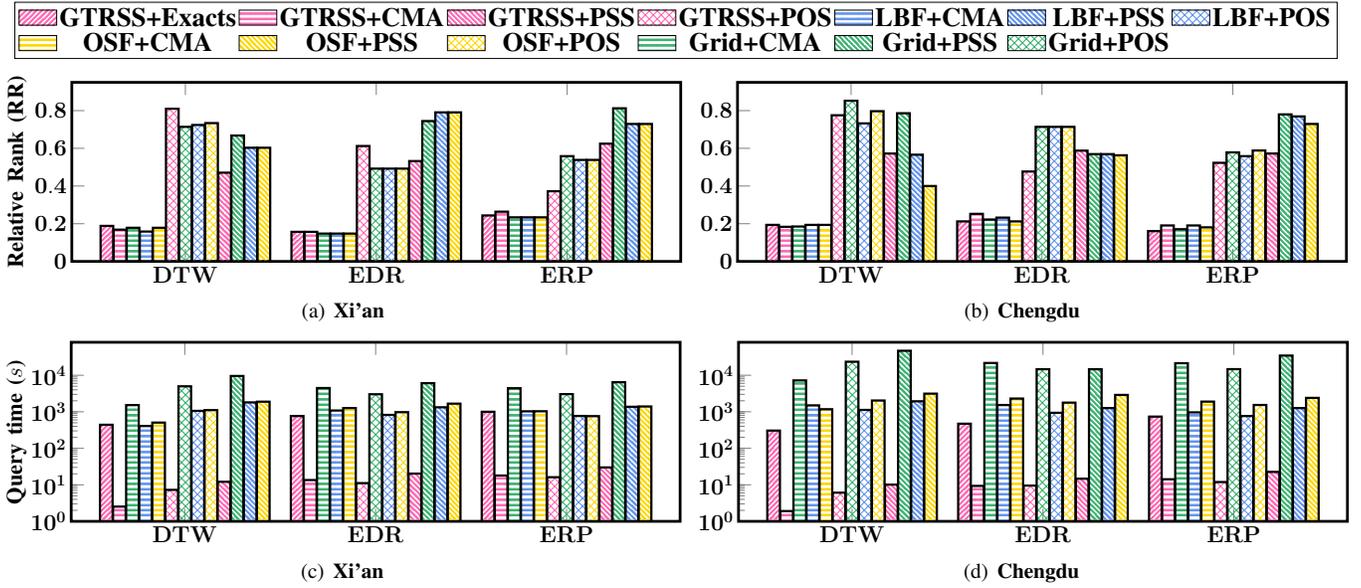

Fig 7: **The performance comparison results of different filtering methods**

TABLE I: **Performance of different methods under popular trajectory similarity measurement.**

| Dataset | Distance Function | DTW | | | EDR | | | ERP | | |
|---|---|---|---|---|---|---|---|---|---|---|
| | Method | HR20 | R10@50 | HR50 | HR20 | R10@50 | HR50 | HR20 | R10@50 | HR50 |
| Xi'an | T3S | 0.0215 | 0.0531 | 0.0569 | 0.0214 | 0.0565 | 0.0557 | 0.0253 | 0.0564 | 0.0567 |
| | TMN | 0.1197 | 0.2692 | 0.2279 | 0.1350 | 0.3155 | 0.2414 | 0.1516 | 0.3447 | 0.2740 |
| | RSSE | 0.2398 | 0.5076 | 0.3727 | 0.4368 | 0.771 | 0.5514 | 0.4500 | 0.7514 | 0.6078 |
| | GTRSS(CMA) | 0.9215 | 0.9255 | 0.8963 | 0.882 | 0.9305 | 0.8691 | 0.9155 | 0.9245 | 0.8865 |
| | GTRSS(POS) | 0.6785 | 0.6695 | 0.6692 | 0.802 | 0.833 | 0.7488 | 0.8435 | 0.8775 | 0.7848 |
| | GTRSS(PSS) | 0.8333 | 0.851 | 0.7745 | 0.7875 | 0.8205 | 0.7510 | 0.7765 | 0.8255 | 0.6921 |
| Chengdu | T3S | 0.0257 | 0.0618 | 0.0628 | 0.0248 | 0.0624 | 0.0622 | 0.0264 | 0.0608 | 0.0627 |
| | TMN | 0.1022 | 0.2286 | 0.1897 | 0.1131 | 0.2558 | 0.1988 | 0.1321 | 0.2932 | 0.2284 |
| | RSSE | 0.3129 | 0.5683 | 0.4301 | 0.3958 | 0.7098 | 0.5032 | 0.4828 | 0.7570 | 0.6219 |
| | GTRSS(CMA) | 0.5482 | 0.7770 | 0.5939 | 0.8522 | 0.8885 | 0.8288 | 0.8338 | 0.8445 | 0.7995 |
| | GTRSS(POS) | 0.6295 | 0.6202 | 0.6195 | 0.5342 | 0.5703 | 0.5042 | 0.7827 | 0.8005 | 0.7205 |
| | GTRSS(PSS) | 0.8061 | 0.8254 | 0.7563 | 0.7355 | 0.7855 | 0.6716 | 0.7252 | 0.7465 | 0.6701 |

**Metrics:** We evaluate GTRSS using three metrics: HR-20, HR-50 (Section V-B), and R10@50. R10@50 measures the overlap between top-50 predictions and actual top-10 scores, quantifying the model's ability to identify relevant similarities. All metrics range from 0 to 1, with higher values indicating better performance.

We compare GTRSS with the above models on retrieving top-$k$ representative similar subtrajectories. All baseline methods first predict similarity scores and then select top-$k$ results. Average results are reported over 200 queries. A Grid filter first preprocesses each query trajectory to retain spatially similar data trajectories, reducing inefficient searches. ExactS then computes ground-truth similarity scores between filtered trajectories and the query, serving as the reference standard.

As shown in Table I, GTRSS outperforms all baselines across all metrics (R10@50, HR20, HR50). Specifically, T3S uses self-attention for intra-trajectory feature extraction but fails to improve downstream similarity score estimation. TMN's matching-based approach surpasses T3S, yet its coarse-grained features limit performance compared to GTRSS. While TMN's matching-based approach surpasses T3S, its coarse-grained feature extraction significantly undermines its performance compared to the GTRSS. Although RSSE based on learning performs better than T3S and TMN, it still falls short in terms of effectiveness and efficiency compared to the GTRSS. The GTRSS performs better than other methods on R10@50, HR20, and HR50, and is currently the best approach.

### E. Hyper-parameter Analysis

**1) Number of neighbors on per node:** As shown in Fig. 8, a small number of neighbors per node reduces the HR10–HR50 performance. This occurs because fewer neighbors weaken trajectory aggregation effects and reduce the likelihood of sufficiently exploring similar subtrajectories. By contrast, increasing the number of neighbor nodes initially improves performance metrics. However, beyond a threshold, further increases yield no improvement because existing neighbors already enable effective top-k retrieval. Additional neighbors introduce useless connections, increasing computational complexity without enhancing performance.

**2) top-$k$ size:** As shown in Fig. 9, the experiment evaluates top-$k$ hit rate (accuracy) across varying $k$ values. The GTRSS-

TABLE II: **Ablation study in Chengdu dataset**

| Algorithm | DTW | | | | | EDR | | | | | ERP | | | | |
|---|---|---|---|---|---|---|---|---|---|---|---|---|---|---|---|
| | HR-10 | HR-20 | HR-30 | HR-40 | HR-50 | HR-10 | HR-20 | HR-30 | HR-40 | HR-50 | HR-10 | HR-20 | HR-30 | HR-40 | HR-50 |
| GTRSS-GARI | 0.370 | 0.388 | 0.392 | 0.389 | 0.391 | 0.530 | 0.548 | 0.552 | 0.570 | 0.572 | 0.685 | 0.710 | 0.697 | 0.671 | 0.657 |
| GTRSS-Random˙node | 0.755 | 0.778 | 0.753 | 0.739 | 0.724 | 0.855 | 0.863 | 0.843 | 0.858 | 0.838 | 0.830 | 0.830 | 0.793 | 0.776 | 0.746 |
| GTRSS-Record˙inter-results | 0.625 | 0.673 | 0.663 | 0.646 | 0.641 | 0.690 | 0.720 | 0.725 | 0.739 | 0.726 | 0.605 | 0.655 | 0.623 | 0.598 | 0.570 |
| **GTRSS** | **0.800** | **0.795** | **0.783** | **0.771** | **0.759** | **0.870** | **0.890** | **0.880** | **0.898** | **0.878** | **0.870** | **0.875** | **0.843** | **0.815** | **0.781** |

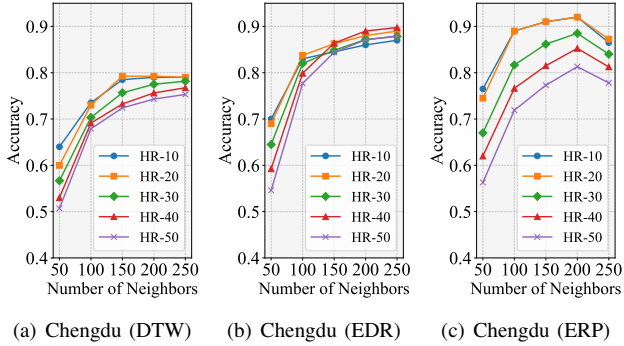

Fig 8: **Impact of Number of Neighbors on Accuracy**

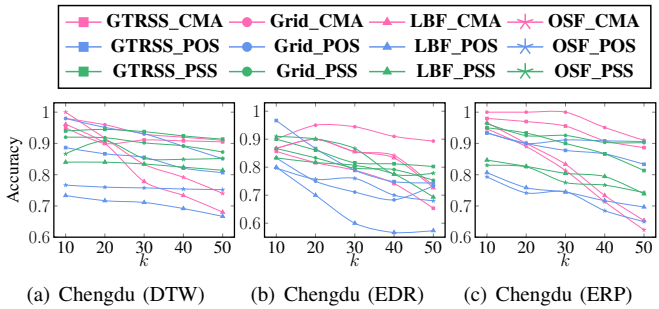

Fig 9: **Impact of $k$ value on accuracy**

based graph index consistently outperforms baseline methods. However, as $k$ increases, accuracy declines due to the limited neighbor count. A small neighbor count may restrict trajectory aggregation and limit result availability for a larger $k$.

*F. Ablation experiments*

We conduct ablation experiments to evaluate the contributions of key components in GTRSS, including GARI, random neighbor nodes, and node record tracking. They play important roles in optimizing HR10–HR50 performance, avoiding local optima, and balancing search efficiency with accuracy.

**GARI Integration** As shown in Table II, GARI substantially improves HR10–HR50 by identifying the most similar trajectory. It accelerates search efficiency in the CNDI and enhances accuracy by avoiding random starting nodes.

**Random Neighbor Nodes** Random neighbor nodes prevents the search from local optima. Without them, metrics degrade due to limited exploration. Conversely, excessive random neighbors weaken clustering effects, reducing performance.

**Node Record Tracking** Recording nodes involved in computations and selecting the final top-k from them improves results. Table II confirms this, showing HR10–HR50 gains from node tracking.

*G. Scalability*

While OSF, Grid, and LBF face significant challenges in terms of scalability when dealing with large trajectory datasets, GTRSS demonstrates superior performance as shown Fig 10, OSF and Grid exhibit a linear increase in query time as the dataset size grows, making them less feasible for large-scale applications. LBF, although more efficient than OSF and Grid, still experiences a notable increase in query time with larger datasets. In contrast, GTRSS leverages advanced indexing and filtering techniques to maintain high efficiency and precision, even as the dataset size increases exponentially. This makes GTRSS a robust and scalable solution for TRSSQ in large-scale data environments.

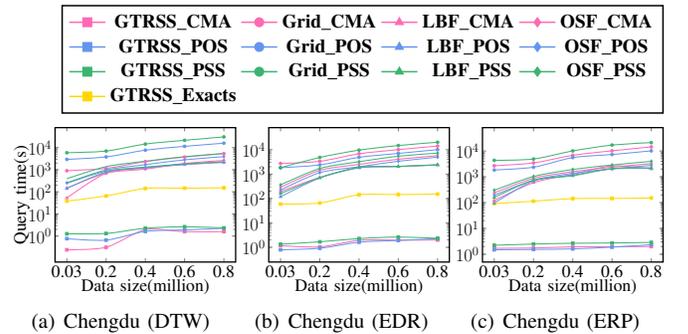

Fig 10: **Scalability comparison across different data sizes**

## VI. CONCLUSION

In this paper, we present a pioneering framework, GTRSS, which utilizes a dual graph index structure to efficiently address the TRSSQ problem. It constructs an upper-level graph on a filtered dataset and a lower-level graph on the whole dataset. A two-step searching mechanism avoids exhaustive dataset processing and significantly enhances query efficiency. To determine the connectivity of nodes in the graphs, we propose the DTSM algorithm to evaluate the similarity between trajectories within the GTRSS framework. To mitigate computational complexity, we incorporate an R-tree filter along with sophisticated pruning techniques in DTSM. Additionally, we record nodes involved in the search process during the online search phase to refine result accuracy. Extensive experiments conducted on real-world datasets validate the efficacy of our framework on the TRSSQ problem and demonstrate the potential applicability of our proposed method in large-scale data environments.